\documentclass[journal,twoside,a4paper]{IEEEtran} 
\usepackage{color}
\makeatletter
\def\blfootnote{\gdef\@thefnmark{}\@footnotetext}
\makeatother
\RequirePackage{filecontents}
\usepackage{cite}
\usepackage{graphicx}
\usepackage{paralist}
 \usepackage{multirow}
 \usepackage[table,xcdraw]{xcolor}
\usepackage[acronym]{glossaries}
\usepackage{amsmath,stackrel}
\usepackage{amsfonts}
\usepackage{mathalfa}
\usepackage{amssymb}
\usepackage{amscd}
\usepackage{tikz}
\usepackage{verbatim}
\usepackage{enumerate}                     
\usepackage{amsthm}
\usepackage{eucal}
\usepackage{psfrag}
\usepackage{algorithmic}
\usepackage{algorithm}
\usepackage{acronym}
\usepackage[normalem]{ulem}

\renewenvironment{IEEEbiography}[1] 
{\IEEEbiographynophoto{#1}}
{\endIEEEbiographynophoto}
\def\W{{\bf W}}
\def\M{{\bf M}}
\def\nmin{s}
\def\nmax{m}
\def\reggamma{P}
\def\Fj{\reggamma}

\def\Fle{F_{\text{$\ell$RIC}}(x)}
\def\Fue{F_{\text{uRIC}}(x)}
\def\Fmine[#1]{F_{\min}(#1)}
\def\Fmaxe[#1]{F_{\max}(#1)}
\def\ric{\delta_s({\bf A})}
\def\rict{{\delta}}
\def\srict{\delta_{s}^{*}(m,n,\epsilon)}
\def\lric{\underline{\delta}_{s}({\bf A})}
\def\lrick1{\underline{\delta}_{{k}_{1}s}({\bf A})}
\def\lrictwos{\underline{\delta}_{2s}({\bf A})}
\def\lrictk1{{\underline{\delta}_{{k}_{1}s}^{*}}}
\def\lrict{{\underline{\delta}_{s}^{*}}(m,n,\epsilon)}
\def\uric{\overline{\delta}_{s}({\bf A})}
\def\urick2{\overline{\delta}_{{k}_{2}s}({\bf A})}
\def\urict{{\overline{\delta}_{s}^{*}}(m,n,\epsilon)}
\def\urictk2{{\overline{\delta}_{{k}_{2}s}^{*}}}
\def\urictwos{\overline{\delta}_{2s}({\bf A})}
\def\maxs{s^{*}}
\def\ppr{P_{\text{PR}}}

\def\rhoa{\rho}

\def\tw{\Psi_{\text{\tiny{TW}}\tiny{1}}}

\theoremstyle{plain}

\theoremstyle{plain}
\newtheorem{theorem}{Theorem}
\theoremstyle{plain}

\theoremstyle{plain}

\theoremstyle{remark}

\theoremstyle{remark}

\theoremstyle{definition}
\newtheorem{definition}{Definition}

\acrodef{maxs}[MSO]{maximum sparsity order}
\acrodef{cs}[CS]{compressed sensing}
\acrodefplural{cs}[CS]{Compressed sensing}
\acrodef{rip}[RIP]{restricted isometry property}
\acrodef{ric}[RIC]{restricted isometry constant}
\acrodef{arip}[ARIP]{asymmetric RIP}
\acrodef{aric}[ARIC]{asymmetric RIC}
\acrodef{rict}[RICt]{RIC threshold}
\acrodef{lric}[LRIC]{lower RIC}
\acrodef{lrict}[LRICt]{LRIC threshold}
\acrodef{uric}[URIC]{upper RIC}
\acrodef{urict}[URICt]{URIC threshold}
\acrodef{rq}[RQ]{Raleigh quotient}
\acrodef{omp}[OMP]{orthogonal matching pursuit}
\acrodef{iht}[IHT]{iterative hard thresholding}
\acrodef{cosamp}[CoSaMP]{compressive sampling matching pursuit}
\acrodef{sp}[SP]{subspace pursuit}
\acrodef{r.v.}[r.v.]{random variable}
\acrodef{pdf}[PDF]{probability density function}
\acrodef{cdf}[CDF]{cumulative distribution function}
\acrodef{ccdf}[CCDF]{complementary CDF}
\acrodef{wm}[WM]{Wishart matrix}
\acrodefplural{wm}[WM]{Wishart matrices}
\acrodef{i.i.d.}{independent, identically distributed}
\acrodef{EED}[EED]{exact eigenvalues distribution}
\acrodef{TW}{Tracy-Widom}
\acrodef{TWDT}{TW Distribution Tail}
\begin{document}
\title{Limits on Sparse Data Acquisition: RIC Analysis of Finite Gaussian Matrices }
\author{Ahmed~Elzanaty,~\IEEEmembership{Student~Member,~IEEE}, Andrea~Giorgetti,~\IEEEmembership{Senior~Member,~IEEE}, and Marco~Chiani,~\IEEEmembership{Fellow,~IEEE} \vspace{-0.2cm}
	\thanks{This work was supported in part by the European Commission under the EuroCPS project and the EU-METALIC II project, within the framework of Erasmus Mundus Action 2.
	This paper was presented in part at the IEEE Statistical Signal Processing Workshop (SSP),  Spain, June 2016.
}
	\thanks{The authors are with the 
		DEI, University of Bologna,
		Via Venezia 52, 47521 Cesena, ITALY
		(e-mail: \{ahmed.elzanaty, andrea.giorgetti, marco.chiani\}@unibo.it).}%
	\thanks{Copyright (c) 2017 IEEE. Personal use of this material is permitted.  However, permission to use this material for any other purposes must be obtained from the IEEE by sending a request to pubs-permissions@ieee.org.}
}
\markboth{ACCEPTED FOR PUBLICATION in IEEE Trans. Inf. Theory}{Elzanaty \MakeLowercase{\textit{et al.}}: {Limits on Sparse Data Acquisition: RIC Analysis of Finite Gaussian Matrices}}
\maketitle
\begin{abstract}
	One of the key issues in the acquisition of sparse data by means of \ac{cs} is the  design of the measurement matrix. Gaussian matrices have been proven to be information-theoretically optimal in terms of minimizing the required number of measurements for sparse recovery. In this paper we provide a new approach for the analysis of the \ac{ric} of finite dimensional Gaussian measurement matrices. The proposed method relies on the exact distributions of the extreme eigenvalues for Wishart matrices.  First, we derive the probability that the \acl{rip} is satisfied for a given sufficient recovery condition on the \ac{ric}, and  propose a probabilistic framework to study both the symmetric and asymmetric \acp{ric}. Then, we analyze the recovery of compressible signals in noise through the statistical characterization of stability and robustness.  The presented framework determines limits on various sparse recovery algorithms for finite size problems. In particular, it provides a tight lower bound on the maximum sparsity order of the acquired data allowing signal recovery with a given target probability. Also, we  derive simple approximations for the \acp{ric} based on the Tracy-Widom distribution.
	\acresetall
\end{abstract}
\begin{IEEEkeywords}
Data acquisition, compressed sensing, restricted isometry property,  Wishart matrices, Gaussian measurement matrices, sparse reconstruction, robust recovery.
\end{IEEEkeywords}
\section{Introduction}
\acp{cs} is an acquisition technique for efficiently recovering a signal from a small set of linear measurements, provided that the sensed data is sparse, i.e.,  the number of its non-zero elements, $s$, is much less than its dimension $n$. If properly chosen, the number of measurements, $m$, can be much smaller than the signal dimension \cite{CanTao:05,Don:06,Don:06a,CanTao:04,CanWak:08,EldKupBol:10,EldKut:12}. 

\acp{cs} based techniques have been exploited to provide efficient solutions for several  problems in signal processing and communication, e.g., source and channel coding, cryptography, random access,  radar, channel estimation, and sub-Nyquist data acquisition \cite{ColRouMag:14,ValColMag:15,SchBocDek:15,KipReeEldGol:17,MotDelRod:17,PotErtParCet:10,DorRau:17,TauHlaEiwRau:10,MisEld:10}. The usability of such applications depends on the maximum sparsity order $s$ such that  recovery is guaranteed with high probability for  given $m$ and $n$. 

{The three main possible approaches to find the maximum sparsity order $s$ guaranteeing recovery of all sparse vectors are based on the \ac{rip} analysis, geometric methods, and coherence analysis.} The \ac{rip} tells how well a linear transformation preserves distances between sparse vectors, and is quantified by the so-called \ac{ric} \cite{CanTao:05}.  In general, the smaller the \ac{ric}, the closer the transformation to an isometry (a precise definition of the \ac{ric} is given later). Geometric based methods are useful for the recovery analysis of exactly sparse signals via $\ell_1$-minimization in the noiseless case  \cite{DonTan:10,RudRom:08,Bay:15,DonTan:10c}. Sparse reconstruction can also be studied looking at the coherence of the measurement matrix. However, the resulting bounds are too pessimistic compared to  RIP-based bounds \cite[eq. (6.9) and eq. (6.14)]{FouRau:13}. This significant gap justifies preferring the \ac{rip} based analysis, whenever bounding the \acs{ric} is feasible. Furthermore, non-uniform recovery guarantees, like those based on Gaussian widths, provide tight bounds for the reconstruction of {a fixed sparse vector}, in contrast to the \ac{rip} method, which considers the recovery of {all sparse vectors} (uniform recovery) \cite{FouRau:13}, \cite{ChaReParcWil:12}.

Moreover, the \ac{rip} theory is more general compared to the geometric approach, as it also considers the stability  for compressible signals and the robustness to noise, under different measurement matrices, for a wider range of sparse recovery algorithms. In fact, sufficient conditions for exact recovery have been obtained for several algorithms in terms of the \ac{ric} (see, e.g.,  \cite{CanTao:05,Can:08,FouLai:08,Fou:10,CaiWanXu:10,MoSon:11,CaiTonZha:13} for $\ell_1$-minimization, \cite{BluDav:08,Fou:12} for \ac{iht}, and \cite{NeeTro:09,DaiMil:09,NeeVer:10,Zha:11} for greedy algorithms).

It has been shown by using information-theoretic methods that Gaussian random matrices with \ac{i.i.d.} entries are optimal in terms of minimizing the number of measurements required for recovery\cite{WanWaiRam:10}. Hence, precisely analyzing the \ac{rip} of such matrices is important. In fact, Gaussian matrices have been proved to satisfy the \ac{rip} with overwhelming probability \cite{Don:06a,CanTao:05}.  The two main tools adopted for the proof are the concentration of measure inequality for the distribution of the extreme eigenvalues of a Wishart matrix, and the union bound which accounts for all possible signal supports. However, if the aim is to  quantify the maximal allowable sparsity order $s$ for a given number of measurements, the use of the concentration inequalities leads to overly pessimistic  results. In this regard, in \cite{BlaCarTan:11,BahTan:10,BBahTan:14} an improved analysis was presented, by bounding the asymptotic behavior of the distributions given in \cite{Ede:88} for the extreme eigenvalues of a Wishart matrix instead of the concentration inequalities. Explicit bounds for the \ac{ric} have been obtained in some specific asymptotic regions \cite{BBahTan:14}, but no bounds are known in the general non-asymptotic setting. In fact, for finite measurement matrices the  asymptotic analysis of the eigenvalues in\cite{BlaCarTan:11,BahTan:10,BBahTan:14} gives approximations of the true distributions; therefore, they cannot provide guaranteed bounds for a particular problem dimension ($s,m,n$).


{This paper  provides an accurate statistical analysis  of the \ac{ric} for finite dimensional Gaussian measurement matrices,  supporting the design of    real \ac{cs} applications (involving always finite size problems), with guaranteed recovery probability. } 
In particular, we calculate the tightest, to our knowledge, lower bound on the probability of satisfying the  \ac{rip} for an arbitrary condition on the \ac{ric}. For a specified number of  measurements, the maximal sparsity order can then be found such that perfect recovery is feasible for all $s$-sparse vectors, i.e., the matrix satisfies the \ac{rip},  considering, on a random draw of the measurement matrix, a target probability $1-\epsilon$ of successful recovery. Differently,  the usually adopted asymptotic setting considers that  this probability  tends to $1$ (overwhelming probability). 

To get better estimates on the maximal sparsity order, tight lower bounds on the \acp{cdf} of the \acp{aric} are  derived, based on the exact probability that the extreme singular values of a Gaussian submatrix are within a range.  Hence, starting from the derived \acp{cdf}, we can find thresholds, below which the \acp{aric} lie with a predefined probability. These percentiles allow to calculate a lower bound on the maximal recoverable signal sparsity order, using several reconstruction methods, such as $\ell_1$-minimization, greedy, and \ac{iht} algorithms. The new analysis is used in conjunction with the recovery conditions relaxed to asymmetric boundaries, as suggested in \cite{BlaCarTan:11}, to prove exact recovery  for signals with larger sparsity orders. In this regard, we relax the symmetric \ac{ric} based  condition in \cite{CaiTonZha:13} to a weaker asymmetric one.  Additionally, we provide approximations for the \ac{ric} \acp{cdf} based on the \ac{TW} distribution, along with convergence investigation. In comparison with previous literature, the proposed analysis gives, for finite dimensional problems, a better estimation of the signal sparsity allowing guaranteed recovery. 

 The contributions of this paper can be summarized as follows: 
\begin{itemize}
\item Accurate symmetric and asymmetric \ac{ric} analysis for finite dimensional problems, accounting for the exact distribution of finite Gaussian matrices (differently from previous methods based on asymptotic behavior of the distributions or loose concentration of measure bounds).
\item Limits on compressive data acquisition in terms of the maximum achievable sparsity order guaranteeing arbitrary target reconstruction probability (instead of the common overwhelming probability approach) via various recovery algorithms.
\item Accurate study for stable and robust recovery of compressible signals with tight bounds on the reconstruction error.   
\item Simple approximations for the \acp{ric} based on the \ac{TW} laws. 
\end{itemize}


Throughout this paper, we indicate with $\det(\cdot)$ the determinant of a matrix, with $\mathrm{card}(\cdot)$ the cardinality of a set, with $\|\cdot \|_{q}=(\sum_{i=1}^{n}{|x_{i}|^q)^\frac{1}{q}}$ the $\ell_q$ norm of an $n$-dimensional vector, with $\|\cdot \|$ the $\ell_2$ norm, with $\Gamma(\cdot)$ the gamma function, with $\gamma\left(a; x, y\right)=\int_{x}^{y} t^{a-1} e^{-t} dt$ the generalized incomplete gamma function, with  $\reggamma(a,x)=\frac{1}{\Gamma(a)}\gamma(a;0,x)$ the regularized lower incomplete gamma function, with $P(a;x,y)=\frac{1}{\Gamma(a)} \int_{x}^{y} t^{a-1} e^{-t} dt=\reggamma(a,y)-\reggamma(a,x)$ the generalized regularized incomplete gamma function, with $\mathcal{N}(\mu,\sigma^{2})$ the Gaussian distribution with mean $\mu$ and variance $\sigma^{2}$. 

\section{Mathematical Background}\label{sec.math}
Compressed sensing allows recovering a signal from a small number of linear measurements, under some constraints on both the sensed signal and the sensing system. More precisely, assume that we have
\begin{equation}
{\bf y}={\bf A}{\bf x}
\label{eq.yeqax}
\end{equation}
where ${\bf y} \in {\mathbb{R}}^{m}$ and ${\bf A} \in \mathbb{R}^{m \times n}$ are known, the number of equations is $m < n$, and ${\bf x} \in {\mathbb{R}}^{n}$ is the unknown. Since $m < n$, we can think of ${\bf y}$ as a compressed version of ${\bf x}$. Without other constraints, the system is underdetermined, and there are infinitely many distinct solutions of \eqref{eq.yeqax}. If we assume that at most $s<m$ elements of ${\bf x}$ are non-zero (i.e., the vector is $s$-sparse), then there is a unique solution (the right one) to \eqref{eq.yeqax}, provided that all possible submatrices consisting of $2s$ columns of ${\bf A}$ are maximum rank. The solution can be found by solving the following $\ell_0$-minimization  \cite{CanTao:05}
 \begin{equation}
 {\bf \hat{x}}\,=\arg \min {{\left\|    {\bf x} \right\|}_{0}} \:\:\text{subject to}\:\:{\bf y}={\bf A} {\bf x}
 \label{eq.l0min}
 \end{equation} 
 where ${{\left\| {\bf x} \right\|}_{0}}$  is the number of the non-zero elements of ${\bf x}$.
However, even when the maximum rank condition is satisfied, the solution of \eqref{eq.l0min} is computationally prohibitive for dimensions of practical interest. A much easier problem is to find the $\ell_1$-minimization solution. 
It is proved in \cite{CanTao:05}, under some conditions on ${\bf A}$, that the solution provided by the $\ell_1$-minimization
 \begin{equation}
 {\bf \hat{x}}\,=\arg \min {{\left\|    {\bf x} \right\|}_{1}} \:\:\text{subject to}\:\:{\bf y}={\bf A} {\bf x}
 \label{eq.l1min}
 \end{equation} 
is the same as that of \eqref{eq.l0min}. The conditions on $\bf A$ are given in term of the \ac{ric}.
 %
\begin{definition}[The \ac{ric} \cite{CanTao:05}]
\label{def.arip}
The \ac{ric} of order $s$ of ${\bf A}$, $\ric$, is  the smallest constant, larger than zero, such that the inequalities
\begin{equation}
1-\ric\leq \frac{{\| {\bf A}_S \,{\bf c} \|}^2}{{\| {\bf c}\|}^2}\leq 1+\ric
\label{eq.ripinequalities}
\end{equation}
are simultaneously  satisfied for every ${\bf c} \in {\mathbb{R}}^s$ and every  $m \times s$ submatrix ${\bf A}_S$ of ${\bf A}$ with columns indexed by $S \subset \Omega  \triangleq \{1,2,...,n\}$ with $\mathrm{card}(S)=s$. Under this condition, the matrix $\bf A$ is said to satisfy the \ac{rip} of order $s$ with constant $\ric$.
\end{definition}

 Specifically, the importance of the \ac{rip} in \ac{cs} comes from the possibility to use the computationally feasible $\ell_1$-minimization instead of the impractical $\ell_0$ one, under some constraints on the \ac{ric}. For example, it was shown that the $\ell_1$ and the $\ell_0$ solutions are coincident for every $s$-sparse vectors $\bf x$ if $\ric<\rict$ with  $\rict=1/3$ \cite{CaiTonZha:13}.  			

The next question is how to design a matrix ${\bf A}$ with a prescribed RIC. One possible way to design ${\bf A}$ consists simply in randomly generating its entries according to some statistical distribution. In this case, for a given $n$, $s$ and $\rict$,  the target is to find a way to generate ${\bf A}$ such that the probability $\mathbb{P}\left\{\ric <\rict \right\}$ is close to one. An optimal choice is to build the measurement matrix ${\bf A}$ with \ac{i.i.d.} entries $a_{i,j}\sim \mathcal{N}(0,1/m)$\cite{CanTao:05,WanWaiRam:10}. 
Then, in order to find the number of measurements $m$ needed, we start by using the Rayleigh quotient inequality for a fixed $S$ 
 \begin{equation}
{\lambda}_{\min}(\W) \leq \frac{{\| {\bf A}_{S}{\bf \,c} \|}^2}{{\| {\bf c}\|}^2} \leq {\lambda}_{\max}(\W)
\label{eq.RQ}
\end{equation} 
 where $\W={\bf {A}}_{S}^{T} {\bf A}_{S}$, and ${\lambda}_{\min}({\W})$ and ${\lambda}_{\max}({\W})$ are its minimum and maximum eigenvalues, respectively.
 Considering that the inequalities in \eqref{eq.ripinequalities} should be satisfied for all the $s$-column submatrices of ${\bf A}$, the \ac{ric} constant can be written as
\begin{align}\label{eq.sric}
\!\!\!\!\!\ric\!=\! \max\left\{\! 1\!-\!\!\!\!\!\min_{\substack{S \subset \Omega \\ \mathrm{card}(S)\!=\!s}}\!\!\!{{\lambda}_{\min}}(\W),\! \! \max_{\substack{S \subset \Omega \\ \mathrm{card}(S)\!=\!s}}\!\!\!{{\lambda}_{\max}(\W)}\!-\!1 \!\right\}.
\end{align}
Hence, the probability that the measurement matrix satisfies the \ac{rip} with  a \ac{ric} at most $\rict$, denoted as $\beta(\rict) \triangleq \mathbb{P}\left\{\ric \leq \rict   \right\}$, is represented by
\begin{align}
\label{eq.probripsatisfiedunionbound}
\!\!\!\!\!\!\!\beta(\rict) \!\!=\!\! \mathbb{P}\!\left\{\!\min_{\substack{S \subset \Omega \\ \mathrm{card}(\!S\!)=s}}\!\!\!\!{{\lambda}_{\min}}({\W}) \!\!\geq\! 1\!-\!\rict,\!\!\!\!\! \max_{\substack{S \subset \Omega \\ \mathrm{card}(\!S\!)=s}}\!\!\!{{\lambda}_{\max}({\W})} \!\leq\! 1\!+\!\rict \right\}\!.
\end{align}
The union bound gives a lower bound for the probability of  satisfying the \ac{rip}  as
%
\begin{equation}
\beta(\rict) \geq 1-{n \choose s}\bigg[ 1-P_{sw}(\rict) \bigg]
\label{eq.probnonRIP}
\end{equation}
where ${n \choose s}$ is the binomial coefficient and $P_{sw}(\rict)$ is the probability that ${\bf A}_S$ is well conditioned defined as: 
\begin{equation}
P_{sw}(\rict)\triangleq \mathbb{P}\left\{1-\rict \leq \lambda_{\min}({\W}), \lambda_{\max}({\W})  \leq 1+\rict\right\}.
\label{eq.psw.general}
\end{equation} 
The probability $P_{sw}(\rict)$ is of fundamental importance, since it determines the performance of \ac{cs}. In the next section, an approach for exactly calculating \eqref{eq.psw.general} for Gaussian  matrices is proposed.
\section{Eigenvalues Statistics}\label{sec.Eigenvalues Statistics}	
In this section, we start by recalling the known concentration inequality based bound on $1-P_{sw}(\delta)$, which is the approach used in \cite{CanTao:05,Don:06}. Then, an alternative method to find $P_{sw}(\delta)$ for Gaussian measurement  matrices are provided. The proposed technique relies on the exact probability that the eigenvalues of $\W$ are within a predefined interval.

\subsection{Eigenvalues Statistics Based on the Concentration Inequality}
\label{subsec.concentration}
Deviation bounds for the largest and the smallest eigenvalues of the Wishart matrix $\W$ are obtained using the concentration of measure inequality \cite{CanTao:05,Don:06}, as
\begin{equation}\label{eq.concent.max}
\mathbb{P}\left\{ \sqrt{\lambda_{\max}({\W})} \geq 1+\sqrt{s/m}+o(1)+t\right\}\leq e^{-m{t}^{2}/2}
\end{equation}
\noindent and
\begin{equation}\label{eq.concent.min}
\mathbb{P}\left\{\sqrt{\lambda_{\min}({\W})} \leq 1-\sqrt{s/m}+o(1)-t\right\}\leq e^{-m{t}^{2}/2}
\end{equation}
where $t>0$ and $o(1)$ is a small term tending to zero as $m$ increases, which will be neglected in the following. Using the inequality $\mathbb{P}\left\{A^c B^c\right\} \geq 1- \mathbb{P}\left\{A \right\}- \mathbb{P}\left\{B\right\}$ where $A, B$ are arbitrary events, and $A^c, B^c$ are their complements, i.e., the union bound, we get 
\begin{multline} \label{eq.psw.boundcandestao}
P_{sw}(\rict) \geq 1-e^{-\frac{1}{2}m \left[{(-1-\sqrt{s/m}+\sqrt{1+\rict})}^{+}\right]^{2}}\\  
-e^{-\frac{1}{2}m \left[{(1-\sqrt{s/m}-\sqrt{1-\rict})}^{+}\right]^{2}}
\end{multline}
where $(x)^{+}= \max\{0,x\}$.  
%
We will see later that this bound, which we use as a benchmark, is  far from the exact probability. 
\subsection{Exact Eigenvalues Statistics}
\label{sec.exactcdfwishart}
We propose a method to compute exactly the probability that a Wishart matrix is well conditioned, i.e, its eigenvalues are within a predefined limit. The method is based on the following recent  result \cite{Chi:15}.
\begin{theorem}
	\label{th:cdfwishartreal}
	The probability that all non-zero eigenvalues of the real \acl{wm} ${\M}={\bf G}^T_S {\bf G}_S$, where ${\bf G}_S$ is $m \times s$ matrix with entries $g_{i,j} \sim \mathcal{N}(0,1)$, are within the interval $[a,b] \subset [0,\infty)$ is
	\begin{align}
	\label{eq:cdfwishartreal}
	\psi_{ms}(a,b)&=\mathbb{P}\left\{a\leq \lambda_{\min}({\M}) , \lambda_{\max}({\M})\leq b\right\}\nonumber \\
	&= K' \,   \sqrt{\det \left({\bf Q}(a,b)\right)}  
	\end{align}
	with the constant
	$$
	K' = \frac{{\pi}^{{\nmin}^{2}}/2}{2^{\nmin \, \nmax/2} \Gamma_{\nmin}(\nmax/2)\Gamma_{\nmin}(\nmin/2)} \, 2^{\alpha \nmin+\nmin (\nmin+1)/2} \prod_{\ell=1}^{\nmin} \Gamma\left(\alpha+\ell\right) \, 
	$$
	where  $\Gamma_{\nmin}(a)\triangleq \pi^{\nmin(\nmin-1)/4}\prod_{i=1}^{\nmin}\Gamma(a-(i-1)/2)$, and $\alpha=\frac{\nmax-\nmin-1}{2}$.
	%
	In \eqref{eq:cdfwishartreal}, when $\nmin$ is even the elements of the $\nmin \times \nmin$ skew-symmetric matrix ${\bf Q}(a,b)$  are
	\begin{multline}
	\label{eq:aij}
	q_{i,j} =
	\left[\Fj\left(\alpha_j,\frac{b}{2}\right)+\Fj\left(\alpha_j,\frac{a}{2}\right)\right] P\left(\alpha_i;\frac{a}{2},\frac{b}{2}\right) \\ -\frac{2}{\Gamma(\alpha_i)} \int_{a/2}^{b/2} x^{\alpha+i-1} e^{-x} \Fj(\alpha_j, x)\, dx 
	\end{multline}
	for $i,j=1,\ldots,\nmin$, where $\alpha_\ell=\alpha+\ell$. When $\nmin$ is odd, the elements of the $(\nmin+1) \times (\nmin+1)$ skew-symmetric matrix ${\bf Q}(a,b)$ are as in \eqref{eq:aij}, with the additional elements
	\begin{eqnarray}  \label{eq:aijoddwishartreal}
	q_{i,\nmin+1}&=& P\left(\alpha_i;\frac{a}{2},\frac{b}{2}\right)   \qquad i=1, \ldots, \nmin  \nonumber \\
	\label{eq:aijodd1} q_{\nmin+1,j}&=&-q_{j,\nmin+1} \qquad\qquad j=1, \ldots, \nmin \\   
	q_{\nmin+1,\nmin+1}&=& 0  \, . \nonumber
	\end{eqnarray}
	Moreover, the elements $q_{i,j}$ can be computed iteratively, without numerical integration or series expansion \cite[Algorithm~$1$]{Chi:15}.
\end{theorem}
Considering that in our case the entries of  ${\bf A}_{S}$ are distributed as $\mathcal{N}(0,1/m)$, the exact  probability that ${\bf A}_S$ is well conditioned is calculated from Theorem \ref{th:cdfwishartreal} as 
\begin{align}\nonumber
P_{sw}(\rict)&=\mathbb{P}\left\{{\lambda}_{\min}({\W}) \geq 1-\rict,{\lambda}_{\max}({\W}) \leq 1+\rict\right\} \\
&=\psi_{ms}\big(m[1-\rict], m[1+\rict]\big) 
\label{eq.probrip.exact}
\end{align} 
where $\psi_{ms}(a,b)$ can now be computed exactly. The exact expression \eqref{eq.probrip.exact} is computationally easy for moderate matrix dimensions (we used it up to $m=1\cdot10^{5}$ and $s=150$). 

%
%
\begin{figure}[t]
	\centering
	\psfrag{S/m}{\scriptsize $s/m$}
	\psfrag{eigenvalues}{\scriptsize  $\mkern-75mu{\lambda}_{\min}^{*}(m,s,\eta)$,  $\lambda_{\max}^{*}(m,s,\eta)$ }
	\psfrag{m=400}{\scriptsize $m=400$}
	\psfrag{m=300}{\scriptsize $m=300$}
	\psfrag{m=200}{\scriptsize $m=200$}
	\includegraphics[clip,width=0.99\linewidth,height=0.77\linewidth]{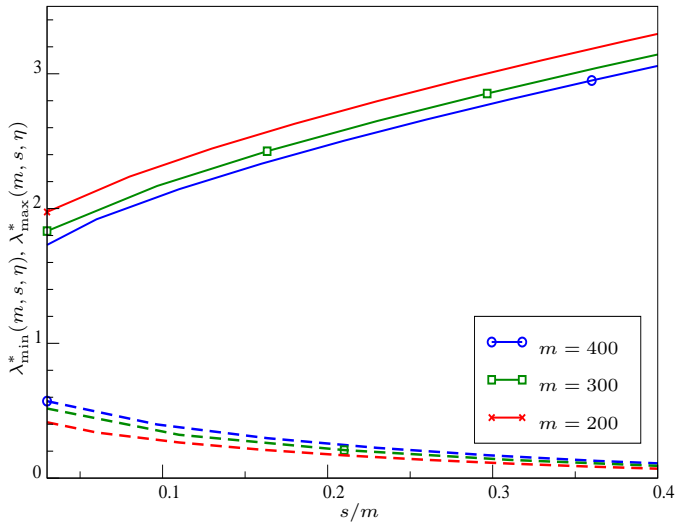}
	\caption{The asymmetric extreme eigenvalues thresholds of the \acl{wm} ${\W}$ as a function of $s/m$, for $\eta =  10^{-10}$. The lower threshold $\lambda_{\min}^{*}(m,s,\eta)$ and the upper threshold $\lambda_{\max}^{*}(m,s,\eta)$ are represented by dashed and solid lines, respectively.}
	\label{Fig.ExtremeEigenvalues}
\end{figure}
%
\subsection{Asymmetric Nature of the Extreme Eigenvalues}\label{subsec.arip}
Clearly, the \ac{ric} in \eqref{eq.sric} depends on the deviation of the extreme eigenvalues from unity. It has been  shown that the smallest and the largest eigenvalues of Wishart matrices asymptotically deviate from $1$\cite{BlaCarTan:11}. Hence, the symmetric \ac{ric}  can not efficiently describe the \ac{rip} of Gaussian matrices. Now, it is essential  to illustrate whether such asymmetric behavior is still valid for finite measurement matrices. 
%
%
%
In this regard, we proposed to find the two  percentiles $\lambda_{\min}^{*}(m,s,\eta)$ and $\lambda_{\max}^{*}(m,s,\eta)$ for the extreme eigenvalues of ${\W}$, such that
\begin{multline*}
\mathbb{P}\left\{\lambda_{\min}({\W})\leq {\lambda}_{\min}^{*}(m,s,\eta) \right\}\\=\mathbb{P}\left\{\lambda_{\max}({\W})\geq {\lambda}_{\max}^{*}(m,s,\eta)\right\} = \eta \,.
\end{multline*}
In fact, such percentiles can be calculated form  the exact eigenvalues distribution in Theorem \ref{th:cdfwishartreal} as
\begin{equation}\label{eq:thrP}
{\lambda}_{\min}^{*}(m,s,\eta)={\psi}_{\min}^{-1}(1-\eta),\,\,\, {\lambda}_{\max}^{*}(m,s,\eta)={\psi}_{\max}^{-1}(1-\eta)
\end{equation}
where ${\psi}_{\min}^{-1}(y)$ and ${\psi}_{\max}^{-1}(y)$ are the inverse of ${\psi}_{ms}(m\,x,\infty)$ and ${\psi}_{ms}(0,m\,x)$, respectively.

In Fig.~\ref{Fig.ExtremeEigenvalues} we report the thresholds ${\lambda}_{\min}^{*}(m,s,\eta)$ and ${\lambda}_{\max}^{*}(m,s,\eta)$ as a function of $s/m$, for some finite values of $m$ and a fixed exceeding probability $\eta = 10^{-10}$. We can see that they asymmetrically deviate from unity, as already observed for asymptotic large matrices in \cite{BlaCarTan:11}. Additionally, since for small values of $m$ the deviation of the extreme eigenvalues from unity is more significant, i.e., the \ac{ric} should be larger, the asymptotic tail behavior of the   eigenvalues distributions in \cite{BlaCarTan:11,BahTan:10,BBahTan:14} cannot be used for upper bounding  the  \acp{ric} in the finite case. 
\begin{definition}[\ac{aric} \cite{FouLai:08,BlaCarTan:11}]\label{def.aric}
	
	The \ac{lric} of order $s$ of ${\bf A}$, $\lric$, is defined as the smallest constant larger than zero that satisfies
	\begin{equation}
	1-\lric \!\leq \!\frac{{\| {\bf A}_{S}\,{\bf c} \|}^2}{{\| {\bf c}\|}^2} \quad \forall\,{\bf c} \in \mathbb{R}^{s},\! \forall\, {S} \subset {\Omega} \!:\!\mathrm{card}(S)\!=\!s
	\label{eq.lric}
	\end{equation}
	and the \ac{uric} of order $s$ of ${\bf A}$,  $\uric$, is defined as the smallest constant larger than zero that satisfies
	\begin{equation}
	\frac{{\| {\bf A}_{S}{\bf \,c} \|}^2}{{\| {\bf c}\|}^2} \! \leq \! 1+\uric \,\,\,\,\, \forall\, {\bf c} \in \mathbb{R}^s,\! \forall\, S \subset \Omega \!:\!\mathrm{card}(S)\!=\!s.
	\label{eq.uric}
	\end{equation}
	%
\end{definition}
Clearly, the relation with the symmetric \ac{ric} is $\ric= \max\{\lric, \uric\}$. 
Moreover, from Definition~\ref{def.aric} and \eqref{eq.RQ}, we can represent the \acp{aric} as
%
\begin{align}
&\lric=1-{\min_{\substack{S \subset \Omega \\ \mathrm{card}(S)=s}}{{\lambda}_{\min}}({\W})}  \label{eq.ldelta.lambda}\\ 
 &\uric=\max_{\substack{S \subset \Omega \\ \mathrm{card}(S)=s}}{{\lambda}_{\max}({\W})}-1.
 \label{eq.udelta.lambda}
\end{align}
%
\section{Symmetric and Asymmetric \acp{ric}}\label{sec.arictheorems} 
The symmetric and asymmetric \acp{ric} of a Gaussian matrix can be seen as functions of the extreme eigenvalues of Wishart matrices as in \eqref{eq.ldelta.lambda} and \eqref{eq.udelta.lambda}, and hence are themselves \acp{r.v.}. 
In this section, we  derive at first lower bounds on the probability of satisfying  \ac{rip} for finite dimensional Gaussian random matrices using the exact eigenvalues distribution, and then a lower bound on the \ac{ric}. Additionally, the \acp{cdf} of the \acp{aric} are lower bounded using the  \acp{cdf} of the extreme eigenvalues. Finally, thresholds for  \acp{aric} that are not exceeded with a target probability are deduced. 

In the following, the analysis derived starting from the exact eigenvalues statistic \eqref{eq.probrip.exact} will be referred as the \ac{EED} based approach. 
\subsection{\ac{rip} Analysis for Gaussian Matrices} \label{sec.symmrip}
A Gaussian matrix is said to satisfy the \ac{rip} of order $s$ if its \ac{ric}, $\ric$, is less than a constant $\rict$ with high probability on a random draw of $\bf{A}$. In other words, if a sufficient condition for perfect reconstruction using a sparse recovery algorithm is satisfied with high probability. This probability can be lower bounded from \eqref{eq.probnonRIP} and \eqref{eq.probrip.exact} as
\begin{align}\label{eq.ripprobexactapprox}
\!\!\beta(\rict, m, n, s)&\!\geq 1\!-{n \choose s}\!\left[1-\vphantom{\tilde{{\psi}_{ms}}}{\psi}_{ms}\big(m[1\!-\!\rict], m[1\!+\!\rict]\big)\right].
\end{align}
The expression \eqref{eq.ripprobexactapprox} gives, to the best of our knowledge, the tightest lower bound on the probability of satisfying the \ac{rip}, $\beta(\rict)$, for finite dimensional Gaussian matrices. This is attributed to  employing the exact joint distribution of the extreme eigenvalues of Wishart matrices, 
providing a quantitatively sharper estimates compared to the concentration bound and the asymptotic approaches. 

When applying \ac{cs}, it is important to estimate the \ac{ric} to assess the recovery property of the measurement matrix. 
Let us define $\delta_{s,\min}^{*}(m,n,\epsilon)$ as the \ac{ric} 
which is exceeded with probability $\epsilon$, such that
\begin{align}\label{eq.sricexactapproxCDF}
\mathbb{P}\{\ric \leq \delta_{s,\min}^{*}(m,n,\epsilon) \} = 1-\epsilon \,.
\end{align}
%
Using \eqref{eq.ripprobexactapprox}  we can upper bound this value as
\begin{equation}\label{eq.sricexactapprox}
\delta_{s,\min}^{*}(m,n,\epsilon) \leq \srict  \triangleq \psi_{ms}^{-1}\left(1-{\epsilon}/{{n \choose s}}\right) 
\end{equation}
where $\psi_{ms}^{-1}(y)$ is the  inverse of $\psi_{ms}\big(m(1-x), m[1+x]\big)$. 
In the following we will refer to $\srict$ in \eqref{eq.sricexactapprox} as the \ac{rict}, where from \eqref{eq.sricexactapproxCDF} and \eqref{eq.sricexactapprox} we have
\begin{equation}\label{eq.rictcdf}
\mathbb{P}\{\ric \leq \delta_{s}^{*}(m,n,\epsilon) \} \geq  1-\epsilon \,.
\end{equation}
%
\subsection{Asymmetric \ac{rip} Analysis for Gaussian Matrices}\label{sec.arics}
%
%
Let $\lric$ be the \ac{lric} as defined in \eqref{eq.ldelta.lambda}. The  \ac{cdf} of the  \ac{lric}, $\Fle$, 
 is lower bounded as
\begin{equation}\label{eq.exactcdflric}
\mathbb{P}\left\{\lric \leq x \right\}\geq 1-{n \choose s}\left[1-\vphantom{\tilde{\psi_{ms}}}\psi_{ms}\big(m[1-x],\infty\big)\right] 
\end{equation}
%
%
%
In fact, from \eqref{eq.ldelta.lambda} the \ac{cdf} of the \ac{lric} $\lric$ is  
\begin{align}
\Fle&=\mathbb{P}\left\{ 1-\min_{\substack{S \subset \Omega \\ \mathrm{card}(S)=s}} {{\lambda}_{\min}}({\W})  \,\, \leq x  \right\} \nonumber \\  
&\geq 1-{n \choose s}\mathbb{P}\left\{{{\lambda}_{\min}}({\W}) \leq 1-x  \right\} \label{eq.lric.cdf}\\  \nonumber
& = 1-{n \choose s}\left[1-\vphantom{\tilde{\psi_{ms}}}\psi_{ms}\big(m\,[1-x],\infty\big)  \right]. 
\end{align}
%
%

Let us define $\underline{\delta}_{s,\min}^{*}(m,n,\epsilon)$ as the \ac{lric} 
which is exceeded with probability $\epsilon$, such that
%
\begin{equation*}
\mathbb{P}\{\lric \leq \underline{\delta}_{s,\min}^{*}(m,n,\epsilon)\}= 1-\epsilon \,.
\label{eq.plricte}
\end{equation*}
This quantity is upper bounded as follows
\begin{align}\label{eq.lricte}
&\!\underline{\delta}_{s,\min}^{*}(m,n,\epsilon) \!\leq\! \lrict \!=\!\psi_{ms,\text{lower}}^{-1}\left(1-{\epsilon}/{{n \choose s}}\right) 
\end{align}
where $\psi_{ms, \text{lower}}^{-1}(y)$ is the  inverse of $\psi_{ms}\big(m[1-x],\infty\big)$. 
In the following we will refer to $\lrict$ as the \ac{lrict}.

Similarly, for the \ac{cdf} of the \ac{uric}, $ \Fue$, we have 
\begin{align}\label{eq.cdfuric}
 \mathbb{P}\left\{\uric \leq x \right\} &\geq 1-{n \choose s}\,\mathbb{P}\left\{\lambda_{\max}(\W)\geq 1+x \right\} \\
 &= 1-{n \choose s}\left[1-\vphantom{\tilde{\psi_{ms}}}\psi_{ms}\big(0, m\,[1+x]\big)  \right]. \nonumber  
\end{align}
Then, we can compute a threshold such that ${\mathbb{P}\{\uric \leq \overline{\delta}_{s,\min}^{*}(m,n,\epsilon)\}= 1-\epsilon}$, which leads to 
\begin{align}\label{eq.uricte}
&\overline{\delta}_{s,\min}^{*}(m,n,\epsilon) \leq \urict\!=\!\psi_{\!ms, \text{upper}}^{-1}\left(\!1\!-\!\frac{\epsilon}{{n \choose s}}\!\right) 
\end{align}
where $\psi_{ms, \text{upper}}^{-1}(y)$ is the  inverse of $\psi_{ms}\big(0,m [1+x]\big)$.
  In the following we will refer to $\urict$ as the \ac{urict}. 

Note that, while previously known approaches refer to infinite dimensional matrices, our analysis accounts for the (always finite) true dimensions of the problem.
\section{Conditions For Perfect Recovery}\label{sec.pr}
In this section, the estimated thresholds for the \acp{ric} (both symmetric and asymmetric) of finite matrices are used to quantify the maximum allowed signal sparsity order for  various recovery algorithms.
%
\begin{definition}[The \acl{maxs}]
\label{def.pt}
Let ${\bf A}$ be a random $m \times n$ measurement matrix,  $s$ be the signal sparsity order, and $0<\epsilon <1$ be an arbitrary constant. The \acl{maxs}, $\maxs$, is the value such that every $s$-sparse vector with $s< \maxs$ can be recovered perfectly with probability $\ppr$ at least $1-\epsilon$ on a random draw of ${\bf A}$. Then the maximum oversampling ratio, a finite regime version of the asymptotic phase transition function, is defined as $\maxs/m$. 
\end{definition} 
The \acl{maxs} is used to compare the performance of different  recovery algorithms and their associated sufficient  conditions. As mentioned before, the perfect reconstruction conditions for many sparse recovery algorithms  are stated in terms of the \acp{ric} \cite{CanTao:05,Can:08,Fou:10,CaiWanXu:10,MoSon:11,CaiTonZha:13,BluDav:08,Fou:12,NeeTro:09,DaiMil:09,NeeVer:10,Zha:11}. 
We now exploit these conditions to 
provide a probabilistic framework for the recovery problem. 
\subsection{Symmetric \ac{ric} Based Sparse Recovery}\label{sec.spr}
About the symmetric \ac{ric}, the sufficient condition for perfect signal recovery via $\ell_1$-minimization can be represented in a generic form as $\delta_{k s}({\bf A})<\delta$, where $k$ is a positive integer and $\delta$ is a constant. As a consequence, the probability of perfect recovery can be bounded as 
%
\begin{equation}
\label{eq:pprRIC}
\ppr \geq \mathbb{P}\left\{\delta_{k s}({\bf A})<\delta  \right\}=\beta(\delta,m,n,ks)
\end{equation}
%
with the proposed \eqref{eq.ripprobexactapprox}. Sufficient recovery condition of this class are, e.g.,  $\ric < 1/3$ \cite{CaiTonZha:13}, $\delta_{2s}({\bf A})<0.6246$ \cite{FouRau:13}, etc.

The inverse problem is the calculation of the maximum sparsity order, for a given $m$ and a given $n$, such that the $\ppr$ is at least $1-\epsilon$. For this target we have 
\begin{equation}
\label{eq:pprRICsmax}
s^{*}= \max\left\{s:  \beta(\delta,m,n,ks)\geq 1-\epsilon\right\}.
\end{equation}
%
\subsection{Asymmetric \ac{ric} Based Sparse Recovery} \label{sec.apr}
Although the asymmetric \acp{ric} are less investigated, it is known that the conditions stated in terms of them lead to tighter bounds for the \acl{maxs} \cite{BlaCarTan:11}. This is attributed to the asymmetric behavior of the extreme eigenvalues for Wishart matrices as analyzed in section \ref{subsec.arip}. 

A general class of sufficient recovery conditions based on the \acp{aric} has the form  
\begin{equation}\label{eq:aricsuff}
\mu(s,{\bf A}) \triangleq f\left(\lrick1\,,\urick2\right)<1
\end{equation}  
where $k_{1}$ and $k_{2}$ are arbitrary positive integers and $f\left(\lrick1\,,\urick2\right)$ is a non-decreasing function in both $\lrick1$ and $\urick2$.
In this regard, we propose a generalization of the symmetric \ac{ric} based condition, $\ric < \frac{1}{3}$, to an asymmetric one.{ In particular, it is possible to prove that if the following condition is satisfied
\begin{equation}
\mu_{\text{\tiny{ECG}}}(s,{\bf A}) \triangleq 2\, \lric+\uric <1
\label{eq.ourcond}
\end{equation}
then all $s$-sparse vectors can be recovered perfectly using $\ell_1$-minimization.\footnote{{The proof is obtained by reformulating equations (33) and (34) in \cite{CaiTonZha:13} to account for the asymmetric \acp{ric}.}}
}
 Other sufficient conditions in the form of \eqref{eq:aricsuff} are found in \cite{FouLai:08,BlaTho:10}. For example, it is shown in \cite{FouLai:08} that if 
\begin{equation}
\mu_\text{\tiny{FL}}(s,{\bf A}) \triangleq \frac{1}{4} \left(1+\sqrt{2}\right) \left(\frac{1+\urictwos}{1-\lrictwos}-1\right) < 1
\label{miofoucart}
\end{equation}
and in \cite{BlaTho:10} that if $${\mu_\text{\tiny{BT}}(s,{\bf A}\!) \triangleq\!\underline{\delta}_{2s}(\!{\bf A})+\left[\underline{\delta}_{6s}({\bf A})+\overline{\delta}_{6s}({\bf A})\right]/4 <1}$$ then perfect reconstruction is also guaranteed.
 
Therefore, for random measurement matrices, the probability of perfect recovery by incorporating the \acp{aric} can be bounded as
\begin{equation}\label{eq:pprARIC}
\ppr \geq \mathbb{P}\left\{\mu(s,{\bf A}) <  1\right\}.
\end{equation}
For the design problem of calculating the maximum sparsity order,  by exploiting  the monotonicity of the function $f(\cdot, \cdot)$, we have 
\begin{align}
&\mathbb{P}\left\{\mu(s,{\bf A})\! \leq\! 1 \right\} \!\geq\! \mathbb{P}\left\{ \lrick1 \!\leq\! \lrictk1 ,\urick2 \!\leq\! \urictk2\!\right\}  \nonumber \\
&\phantom{=} \geq 1 -\mathbb{P}\left\{\vphantom{\urictk2} \lrick1 \geq \lrictk1 \right\}- \mathbb{P}\left\{\urick2 \geq \urictk2 \right\} \label{eq.muepsilon}
\end{align}
for any $\lrictk1$ and $\urictk2$  such that $f\left(\lrictk1\,,\urictk2\right) < 1$. Equation \eqref{eq.muepsilon} is due to the union bound,  \eqref{eq.ldelta.lambda}, and  \eqref{eq.udelta.lambda}. Setting the  bound \eqref{eq.muepsilon} to $1-\eta$ and distributing equally the probability on the lower and upper \acp{ric}, we get 
\begin{equation}\label{eq.epsilondelta}
\mathbb{P}\left\{\urick2 \leq \urictk2 \right\}=\mathbb{P}\left\{\vphantom{\urictk2}\lrick1 \leq \lrictk1 \right\} = 1-\frac{\eta}{2} . 
\end{equation}

Finally, the maximum sparsity order $s^{*}$ is the maximum $s$ compatible with $f\left(\lrictk1,\urictk2\right) < 1$, where $\lrictk1\,, \urictk2$ are calculated from \eqref{eq.lricte} and \eqref{eq.uricte} with $\epsilon=\eta/2$ to satisfy \eqref{eq.epsilondelta}. Then, every sparse vector with  $s<\maxs$ can be perfectly recovered with probability at least $1-\eta$ on a random draw of ${\bf A}$.

Although we focused on $\ell_1$-minimization based recovery, the same approach can be used to estimate the \acl{maxs} using greedy or thresholding algorithms. 
For example, sufficient conditions on the \ac{ric} for perfect recovery  using \ac{cosamp}, \ac{omp}, and \ac{iht} are  $\delta_{4s}({\bf A})<0.4782$ ~\cite{FouRau:13}, $\delta_{13s}({\bf A})<0.1666$ ~\cite{Zha:11,FouRau:13}, and $\delta_{3s}({\bf A})<0.5773$~ \cite{Fou:11}, respectively. Additionally, asymmetric \ac{ric} based conditions  have been obtained in \cite{BlaCarTanThom:11} for  \ac{cosamp}, \ac{iht}, and \ac{sp}. For example, $$\mu_\text{\tiny{BCTT}}(s,{\bf A}) \triangleq 2\sqrt{2} \left(\frac{\overline{\delta}_{3s}({\bf A})+\underline{\delta}_{3s}({\bf A})}{2+\overline{\delta}_{3s}({\bf A})-\underline{\delta}_{3s}({\bf A})} \right)   <1$$ is a sufficient condition for perfect recovery using  \ac{iht}~\cite{BlaCarTanThom:11}. 
\section{Robust Recovery of Compressible Signals}\label{sec.robustness}
Up to now, we have studied the case of perfect recovery of sparse data in noiseless setting. However, in practice signals can also be not exactly sparse, but rather compressible, i.e., the data is well approximated by a sparse signal.
 Moreover, noise can be present during the acquisition process.
 
A measure of the discrepancy between  a compressible signal 
and its sparse representation is the $\ell_1$-error of best $s$-term approximation $\sigma_s({\bf x})_{1}$, defined as 
\begin{equation}
\sigma_s({\bf x})_{1} \triangleq \inf\{ {{{\left\| {\bf x}-{\bf x}_s \right\|}_{1}}, \quad {\bf x}_s \in {\mathbb R}^n  \text{ is $s$-sparse}}\}\,.
\end{equation}
Hence, a signal is well approximated by an $s$-sparse vector if $\sigma_s({\bf x})_{1}$ is small~\cite{FouRau:13}. 
%
Besides considering compressible signals, we can also include the measurement noise in the model, so that the measured vector can be written as
\begin{equation}
{\bf y}={\bf A}{\bf x}+{\bf z}
\label{eq.yeqaxnoise}
\end{equation}
where ${\bf z}$ is a bounded noise with ${\left\| {\bf z} \right\|}\leq \kappa$. Assuming $\kappa$ is known, we can account for the noise term by modifying the constraint in the $\ell_1$-minimization problem \eqref{eq.l1min} as
\begin{equation}
\hat{\bf {x}}\,=\arg \min {{\left\|    {\bf x} \right\|}_{1}} \:\:\text{subject to}\:\:{{\left\| {\bf y}-{\bf A} {\bf x} \right\|}}\leq \kappa \,.
\label{eq.bpdn}
\end{equation} 
 This algorithm is called quadratically constrained $\ell_1$-minimization \cite{Rob:96}. 
 There are also other algorithms for sparse recovery in noisy cases, e.g., Dantzig selector \cite{CanTao:07}, basis pursuit denoising \cite{CheDon:94}, denoising-orthogonal approximate message passing \cite{XueMaYua:16}, etc. 
 
 For the model illustrated in \eqref{eq.yeqaxnoise}, we cannot guarantee perfect signal recovery, but rather an approximate reconstruction can be assured with bounded error. For example, it was shown in  \cite{CaiTonZha:13} that if $\ric <1/3$, the error after recovery can be bounded by a weighted combination of $\kappa$ and $\sigma_s({\bf x})_{1}$, i.e.,
 \begin{equation}
{{\left\| \hat{{\bf x}}-{\bf x} \right\|}} \leq {C_1}\kappa+{C_2}\frac{\sigma_s({\bf x})_{1}}{\sqrt{s}}
\label{eq.optimalinstance}
 \end{equation}
where 
\begin{align}
\label{eq:C1star}
C_1\left(\ric\right)&=\frac{\sqrt{8\,\big[1+\ric\big]}}{1-3\,\ric}\\
C_2\left(\ric\right)&=\frac{\sqrt{8}\bigg[2\,\ric+\sqrt{\big[1-3\,\ric\big]\ric}\,\bigg]}{1-3\,\ric}+2\,.
\label{eq:C2star}
\end{align}
The constants $C_{1}$ and $C_{2}$ give an insight about both the robustness (ability to handle noise) and the stability (ability to handle compressible signals) of the recovery algorithm, respectively. 

 When ${\bf A}$ is a random matrix, both $C_{1}$ and $C_{2}$ are random variables. 
To characterize their statistical distribution, we propose to find a bound on the threshold $C_{i,\min}^{*}$, with $i=1,2$, which is not exceeded with a predefined probability $\epsilon_{i}$, i.e., 
 \begin{align}\label{c1min}
 	\mathbb{P}\left\{C_{i}\left(\ric\right)\leq C_{i,\min}^{*} \right\} &= 1-\epsilon_{i}\,. 
 \end{align}
 Noting that  $C_{i}\left(\ric\right)$  is monotonically increasing in $\ric$, we  have %
 \begin{multline}\label{c1cdf}
 \mathbb{P}\left\{C_{i}\left(\ric\right)\leq C_{i}\left( \delta_{s}^{*}(m,n,\epsilon_{i})\right) \right\}\\
 \qquad=\mathbb{P}\left\{\ric \leq  \delta_{s}^{*}(m,n,\epsilon_{i}) \right\} \geq 1-\epsilon_{i}
 \end{multline}
 where the \ac{rict} $\delta_{s}^{*}(m,n,\epsilon_{i})$ can be calculated  from \eqref{eq.sricexactapprox}. 
 Consequently, from \eqref{c1min} and \eqref{c1cdf}  we upper bound  $C_{i,\min}^{*}$  as
 \begin{align}
 	C_{i,\min}^{*} &\leq C_{i}^{*} \triangleq C_{i}\left( \delta_{s}^{*}(m,n,\epsilon_{i})\right).
 \end{align}

 The inverse problem  is finding  the maximum sparsity order, for a given $m$ and a given $n$, such that the \ac{r.v.} $C_{i}$, with $i=1,2$, is less than a targeted constant $c_{i}$ with probability  at least $1-\epsilon_{i}$. For this aim we have 
 \begin{equation*}
 	\label{eq:C1smax}
 	s^{*}= \max\left\{s: C_{i}\left( \delta_{s}^{*}(m,n,\epsilon_{i})\right)  \leq c_{i}  \right\}.
 \end{equation*}

 Analogous results relating the recovery error with  $\sigma_s({\bf x})_{1}$ and $\kappa$  have been obtained for different  algorithms under suitable  symmetric and asymmetric \ac{ric} based sufficient conditions \cite{FouLai:08,SaaChaYil:08,BlaCarTanThom:11,JacHamKenJal:11,ZenSoJia:16}. By following the same approach, the proposed methodology can be applied to describe the statistics of the stability and robustness constants also for these cases.      
\section{Tracy-Widom Based RIC Analysis}\label{sec.gammaapprox}
Although the proposed framework based on the exact distribution of the eigenvalues  \eqref{eq.probrip.exact} provides tight bounds on the \acp{ric}, it could be computationally expensive for large matrices, for which easier  approaches  are preferred. 

In this section, we  derive approximations for the \acp{ric} of finite matrices based on the \ac{TW} distribution, much tighter than those obtained from  concentration of measure inequalities. Also, we study the convergence rate of the distribution of extreme eigenvalues to those based on the \ac{TW} by exploiting the small deviation analysis of the extreme eigenvalues around their mean. In particular, we prove that  \ac{TW} based distributions approximate the eigenvalues statistics of finite Gaussian matrices with exponentially small error in $m$, leading to accurate estimation of the  \acp{ric}.   

In fact, it is well known that the distribution of the smallest and  largest eigenvalues of Wishart  matrices tend, under some conditions, to a properly scaled and shifted \ac{TW} distributions \cite{TraWid:94,Joh:00,Joh:01,TraWid:09,bor:10,ma:12,BasCheZha:12}. Specifically, it has been shown that for the real \acl{wm} ${\M}$ when $m,s {\xrightarrow{}} \infty$ and $m/s {\xrightarrow{}}  \gamma \in (0,\infty)$
\begin{equation}
\frac{\lambda_{\max}({\M})-\mu_{ms}}{\sigma_{ms}}  \xrightarrow[]{\mathcal{D}}  \mathcal{TW}_1
\label{eq.lambdamax.TW1}
\end{equation}
where $\mathcal{TW}_1$ is a  \acl{TW} \ac{r.v.}  of order $1$ with \ac{ccdf} $\tw(t)$, 
$\mu_{ms}={(\sqrt{m}+\sqrt{s})}^{2}$, and ${\sigma}_{ms}=\sqrt{\mu_{ms}}(1/ \sqrt{s}+1/\sqrt{m})^{1/3}$\cite{Joh:01}. More precisely, from the convergence in distribution definition and letting $\rho \triangleq s/m$ we have
\begin{align}\label{eq.limtw}
&\lim\limits_{m\xrightarrow{} \infty} \mathbb{P}\left\{\lambda_{\max}({\M}) \geq  \mu_{ms}+t\,{\sigma}_{ms} \right\}=\nonumber\\
&\lim\limits_{m\xrightarrow{} \infty} \mathbb{P}\left\{\lambda_{\max}({\W}) \geq  (1+\sqrt{\rho})^{2}+t\, m^{-\frac{2}{3}}\,  \rhoa^{-\frac{1}{6}}\left(1\!+\!\sqrt{\rhoa}\right)^{\frac{4}{3}}  \right\} \nonumber\\
&\hphantom{\lim\limits_{m\xrightarrow{} \infty}}=\tw(t).
\end{align}

Similarly, for the smallest eigenvalue, when $m,s {\xrightarrow{}} \infty$ and $m/s {\xrightarrow{}}  \gamma \in (1,\infty)$\cite{ma:12} 
%
\begin{equation}
- \,  \frac{\ln \lambda_{\min}({\M})-v_{ms}}{\tau_{ms}}  \xrightarrow[]{\mathcal{D}} \mathcal{TW}_1
\label{eq.ln.lambdamin.TW1}
\end{equation} 
with scaling and centering parameters 
\begin{equation*}
{\tau}_{ms}=\frac{{\left[(s-1/2)^{-1/2}-(m-1/2)^{-1/2} \right]}^{1/3}}{\sqrt{m-1/2}-\sqrt{s-1/2}}
\label{eq.tau}
\end{equation*}
\begin{equation*}
v_{ms}=2 \ln\left(\sqrt{m-1/2}-\sqrt{s-1/2} \right) + \frac{1}{8}\tau_{ms}^{2}.
\label{eq.v.lambdamin}
\end{equation*}
%

Regarding the RIC analysis for finite Gaussian matrices, let $\urict$, $\lrict$, and $\srict$ be the \aclp{rict} as defined in \eqref{eq.uricte}, \eqref{eq.lricte}, and \eqref{eq.sricexactapprox}, respectively. We will show that they can be approximated as
\begin{align} 
&\urict \simeq \overline{\delta}^{*}_{\text{\tiny{TW}}} \triangleq \,m^{-\frac{2}{3}}\rhoa^{-\frac{1}{6}}\left(1+\sqrt{\rhoa}\right)^{\frac{4}{3}}
\tw^{-1}\left(\epsilon/{n \choose s} \right)\nonumber \\
&\hphantom{\urict < \overline{\delta}^{*}_{\text{\tiny{TW}}} \triangleq}+\rho +2\sqrt{\rho} \label{eq.approxuricte}\\
&\lrict \simeq \underline{\delta}^{*}_{\text{\tiny{TW}}}\triangleq 1\!-\!\frac{1}{m}  \exp\Bigg(\!{v_{ms}- \tau_{ms} \tw^{-1}\bigg(\epsilon/{n \choose s}\bigg)  }\!\Bigg)\label{lricte2}\\
&\srict \simeq {\delta}^{*}_{\text{\tiny{TW}}}\triangleq \widetilde{P}_{sw}^{-1}\Bigg(1-\epsilon/{n \choose s}\Bigg) \label{eq.approxricte}
\end{align}
 for  $\overline{\delta}^{*}_{\text{\tiny{TW}}}$, $\underline{\delta}^{*}_{\text{\tiny{TW}}}$, and ${\delta}^{*}_{\text{\tiny{TW}}}$ less than one,
where $\tw^{-1}(y)$ is the inverse of the \ac{TW}'s \ac{ccdf} and $\widetilde{P}_{sw}^{-1}(y)$ is the inverse of 
\begin{multline}\label{eq.ripprobexactapprox2}
\widetilde{P}_{sw}(x)\triangleq 1- \tw\Bigg(\frac{v_{ms}-\ln \big(m[1-x]\big)}{\tau_{ms}}\Bigg)\\
- \tw\Bigg(\frac{m[1+x]-\mu_{ms}}{\sigma_{ms}}   \Bigg) .
\end{multline}

In order to prove these formulas, at first the convergence rate of the extreme eigenvalue distributions to those based on  the \ac{TW} is provided. For the \ac{uric}, it has been shown in  \cite[Theorem $2$]{Ledoux:10b} that there exists a constant $c>0$, depending only on $\rho$, such that  
\begin{align}\label{eq.ledoux}
&\mathbb{P}\Big\{\lambda_{\max}(\M) \geq \mu_{ms}[1+z]\Big\} \leq c \exp\left(-\frac{1}{c}\,s\,z^{\frac{3}{2}}\right) 
\end{align}   
for all $m> s\geq 1$ and $0<z \leq 1$. This small deviation analysis provides tighter bounds compared to the concentration inequality \eqref{eq.concent.max} and   Edelman bound \cite[Lemma $4.2$]{Ede:88} used for large $m$ in \cite{BlaCarTan:11,BahTan:10,BBahTan:14}. From \eqref{eq.ledoux}, the L.H.S. of \eqref{eq.limtw} can be tightly bounded for finite $m$  and for ${t\leq m^{2/3} \rhoa^{{1}/{6}}\left(1+\sqrt{\rhoa}\right)^{{2}/{3}}}$ as
\begin{multline}
\mathbb{P}\left\{\lambda_{\max}({\W}) \geq  (1+\sqrt{\rho})^{2}+t\, m^{-\frac{2}{3}}\,  \rhoa^{-\frac{1}{6}}\left(1\!+\!\sqrt{\rhoa}\right)^{\frac{4}{3}}  \right\}\\
\leq c \exp\left(-\,c_{1}\, t^{\frac{3}{2}}\right)
\end{multline}
where $c_{1} \triangleq {c}^{-1}\,\rhoa^{{3}/{4}}\left(1\!+\!\sqrt{\rhoa}\right)^{-1}$. Regarding the R.H.S, for sufficiently large $t$ we have
\begin{equation}\label{eq.twtailbound}
\tw(t) \leq c_{2} \exp\left(- c_{3}\, t^{\frac{3}{2}}  \right)
\end{equation}
where $c_{2}>0$ and $c_{3}>0$ are  constants \cite[eq. (2)]{Aub:05}, \cite{DumVir:11}. 
Now the error in using the \ac{TW} can be bounded as
\begin{align}\label{eq.convtwlambdamax}
&\bigg| \mathbb{P}\left\{\lambda_{\max}({\W}) \geq  (1+\sqrt{\rho})^{2}+t\, m^{-\frac{2}{3}}\,  \rhoa^{-\frac{1}{6}}\left(1\!+\!\sqrt{\rhoa}\right)^{\frac{4}{3}}  \right\} \nonumber \\
&\vphantom {\frac{1}{2}}-\tw(t)  \bigg| \leq c_{4} \exp\left(- c_{5}\, t^{\frac{3}{2}} \right)
\end{align}
where $c_{4} = \max\{c,c_{2}\}$ and $c_{5} = \min\{c_{1},c_{3}\}$. Therefore, the error due to approximating  $\mathbb{P}\left\{\lambda_{\max}({\W}) \geq  1+x  \right\}$ in  \eqref{eq.cdfuric} by that of the \ac{TW} can be bounded from  \eqref{eq.convtwlambdamax}  as 
\begin{align}\label{eq.convuricccdf1}
&\bigg| \mathbb{P}\left\{\lambda_{\max}({\W}) \geq  1+x  \right\}-\!\tw \bigg(\left(x\!-\!2\sqrt{\rho}\!-\!\rho \right) \nonumber \\
&\times m^{\frac{2}{3}}\,  \rhoa^{\frac{1}{6}}\left(1\!+\!\sqrt{\rhoa}\right)^{-\frac{4}{3}} \bigg) \bigg| \leq  c_{4} \exp\left(-  m \left(x-2\sqrt{\rho}-\rho \right)^{\frac{3}{2}} \right.    \nonumber \\
& \left. \times c_{5}\, \rhoa^{\frac{1}{4}}\left(1+\sqrt{\rhoa}\right)^{-2}     \right)
\end{align}
for $x\leq 2\left(1\!+\!\sqrt{\rhoa}\right)^{2}\!-\!1$.\footnote{Note that $x \leq 1$ is a stronger condition than $x\leq 2\left(1\!+\!\sqrt{\rhoa}\right)^{2}\!-\!1$.} Hence, the absolute error in approximating the exact probability with that based on the \ac{TW} distribution is exponentially small in $m$ and the \ac{urict} can be approximated by \eqref{eq.approxuricte}.


A similar reasoning can be used to derive the thresholds for the lower and symmetric \acp{ric} (the proof is not reported here for the sake of conciseness). 

Finally, we would like to remark that \acl{TW} based approaches could be used not only for Wishart ensembles, but also for a wider class of matrices like those drawn from some sub-Gaussian distributions, e.g., Rademacher and Bernoulli measurement matrices. This is motivated by the universality of the \ac{TW} laws for the extreme eigenvalues of large random matrices \cite{FelSod:10,Pec:09}, although further research is required to investigate such extensions. 
\section{Numerical Results}\label{sec.results}
In this section, numerical results are presented to compare the proposed  exact and \ac{TW}  approaches with the concentration inequalities, for analyzing the probability that  the \ac{rip} is satisfied. Moreover, the statistics of  the \acp{ric}, the probability of perfect reconstruction, the \acl{maxs} for various recovery algorithms, and the robustness and stability constants are also investigated.

\begin{figure}[!t]
	\psfrag{s/n}{\scriptsize $s/n$}
	\psfrag{ripprob}{\scriptsize \hspace{-0.76cm} $\mathbb{P}\left\{\ric \geq 1/3 \right\}$}
	\psfrag{Upper bound}{\scriptsize Concentr. bound}\centering
	\psfrag{Exact approach}{\scriptsize \ac{EED} approach}\centering
	\psfrag{Gamma approx.}{\scriptsize \ac{TW} approx.}\centering
	\psfrag{10e-0}{1}
	\psfrag{e-4}{\scriptsize $\times 10^{-4}$}
	\includegraphics[clip,width=0.95\linewidth,height=0.75\linewidth]{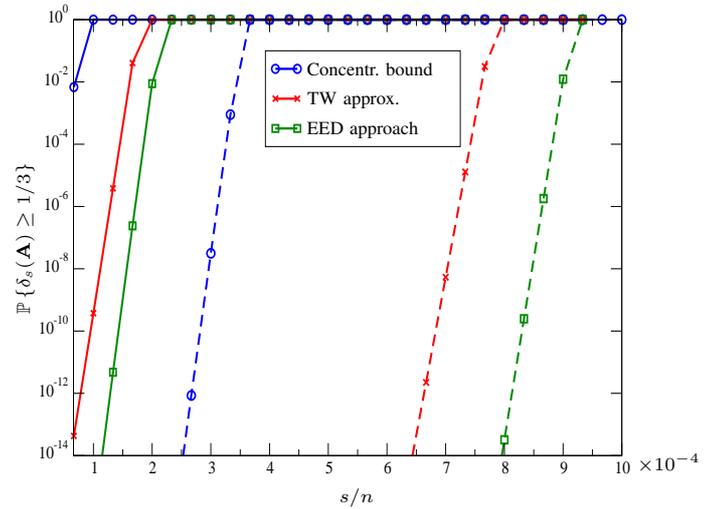}\centering
	\caption{Symmetric \ac{rip}: upper bounds on the probability of not satisfying the \ac{rip}, $\mathbb{P}\left\{\ric \geq 1/3 \right\}$,  for $m/n=0.1$ (solid) and $m/n=0.4$ (dashed). The signal dimension is $n=3\cdot10^4$. Curves obtained through the concentration bound, \eqref{eq.probnonRIP} and \eqref{eq.psw.boundcandestao}, the \ac{EED}, \eqref{eq.ripprobexactapprox}, and the \ac{TW} approximation, \eqref{eq.probnonRIP} and \eqref{eq.ripprobexactapprox2}.}
	\label{fig.RIP_prob2}
\end{figure}

Fig.~\ref{fig.RIP_prob2} shows upper bounds on the probability of not satisfying the \ac{rip}, $\mathbb{P}\!\left\{\ric \!\geq\! 1/3 \right\}$, using the \ac{EED} based approach \eqref{eq.ripprobexactapprox}, the \ac{TW} approximation \eqref{eq.probnonRIP}, \eqref{eq.ripprobexactapprox2}, and the concentration bound \eqref{eq.probnonRIP}, \eqref{eq.psw.boundcandestao}. Note that when the sparsity level is beyond some threshold value, the probability of not satisfying the \ac{rip} rapidly increases from zero to one. 
 This figure also illustrates the limit on the maximum sparsity ratio that still permits satisfying the \ac{rip} with a targeted probability. 
 We can see that the \ac{EED} based approach indicates higher sparsity ratios (less sparse vectors) compared to those estimated by the well-known concentration bound (more than $220\%$ increase in $s/n$ when the probability is $10^{-14}$ and $m/n=0.4$). In fact, the  concentration inequality is quite loose in bounding the probability that a submatrix is ill conditioned, $1-P_{sw}(\rict)$, and consequently in analyzing the \ac{rip}. 

\begin{figure}[!t]
	\psfrag{s/m}{\scriptsize $s/m$}
	\psfrag{m/n}{\scriptsize $m/n$}
	\psfrag{1e-3}{\scriptsize$ \times 10^{-3}$}
	\includegraphics[clip,width=0.99\linewidth,height=0.77\linewidth]{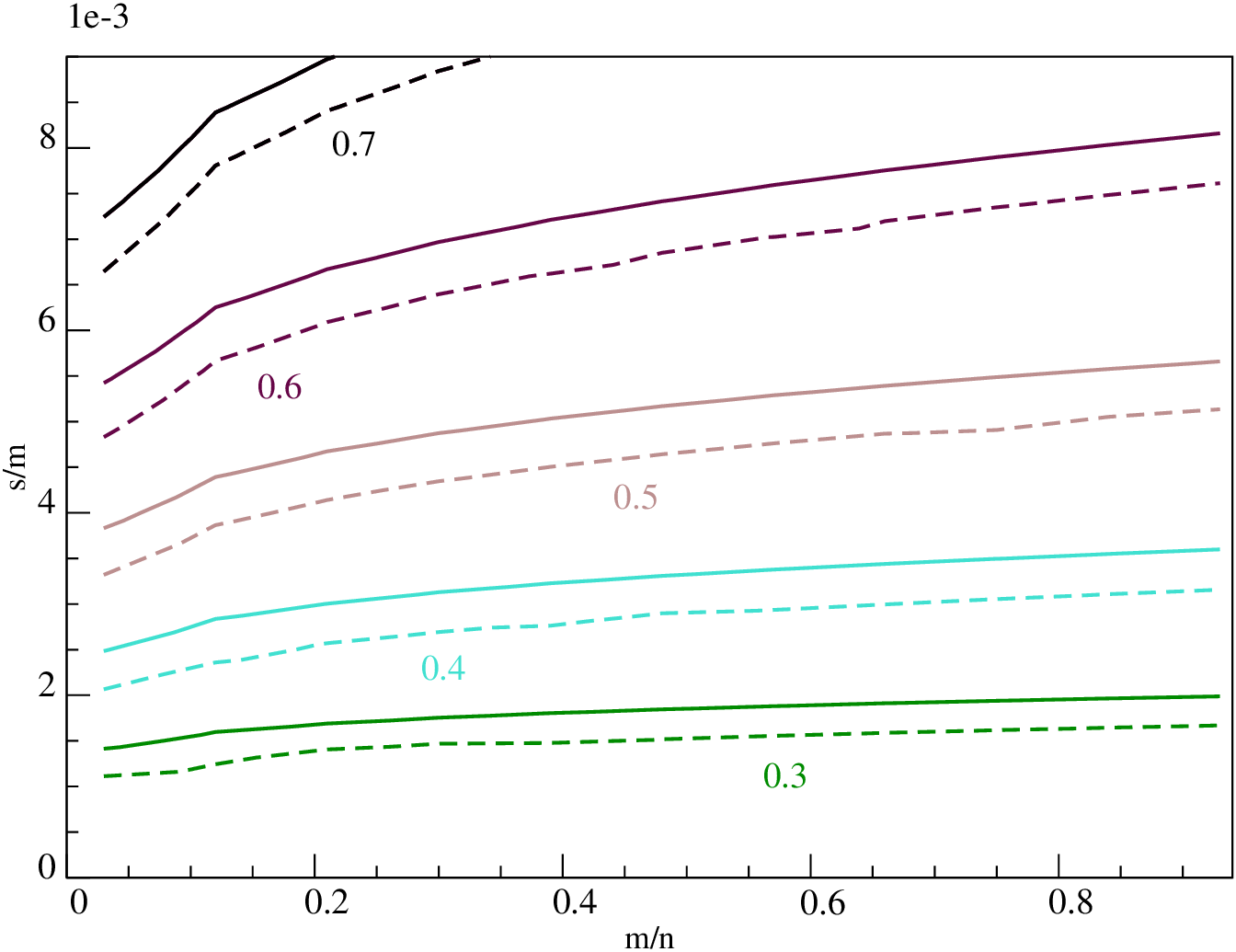}\centering
	\caption{Level sets of the upper \acl{rict} $\urict \in \{0.3,0.4,0.5,0.6,0.7\}$ such that $\mathbb{P}\{\uric \geq \urict\} \leq \epsilon$, using the \ac{EED} (solid) and \ac{TW} (dashed), for $m=4000$ and $\epsilon=10^{-3}$.}
	\label{fig.exact_approx_uric}
\end{figure}

Regarding the \acp{aric}, the upper \ac{ric} thresholds, $\urict$, computed by means of \eqref{eq.uricte} and \eqref{eq.approxuricte}, are plotted in Fig. \ref{fig.exact_approx_uric} for an excess probability $\epsilon=10^{-3}$, as a function of the compression ratio, $m/n$, and the oversampling ratio, $s/m$. In this figure, we set $m=4000$ and vary $n$ from $2\cdot 10^5$ to $4000$. As can be noticed the \ac{TW} approximation is quite accurate. 
%

{To further investigate the \ac{ric} bounds, we report in Table~\ref{tab.rics} both the \ac{lric} and \ac{uric} thresholds for different $m/n$ using various approaches: the proposed \ac{EED} \eqref{eq.lricte}, \eqref{eq.uricte}, the \ac{TW} approximation \eqref{lricte2}, \eqref{eq.approxuricte}, the empirical lower bounds in \cite{DosPeyFad:10}, and the asymptotic bounds in \cite{BlaCarTan:11}, \cite{BahTan:10}. 
  We can see that  the upper bounds  on the \acp{ric} obtained from the \ac{EED} approach  is  sharp, with small differences from the empirical  lower bounds (averaged over $100$ different realizations) indicated by \cite{DosPeyFad:10}.  

\renewcommand{\arraystretch}{1.5}
\begin{table*}[]
	\centering
	\caption{{The \ac{ric} thresholds using the \acs{EED} bound and \ac{TW} approximation for $\epsilon=10^{-2}$, empirical averaged lower bounds \cite{DosPeyFad:10}, BCT \cite{BlaCarTan:11}, and BT \cite{BahTan:10}  approaches, for $m=2000$ and $s=4$. For each $m/n$, the two rows give the upper and lower \ac{ric}.}} 
	\label{tab.rics}
	\begin{tabular}{|ccc|c| c  c |}
		\hline
		\cellcolor[HTML]{FAF8EA}                        & \multicolumn{3}{|c|}{\cellcolor[HTML]{FCF5E4}Finite}                                                                                                            &  \multicolumn{2}{c|}{\cellcolor[HTML]{E9F6E3}Asymptotic}        \\
		\cline{2-6} 
		\multirow{-2}{*}{\cellcolor[HTML]{FAF8EA}$m/n$ $\downarrow$}   & \multicolumn{1}{|c}{\cellcolor[HTML]{FFFEFC}\begin{tabular}[c]{@{}c@{}}\acs{EED}\vspace{-2mm}\\\vspace{-1.5mm} upper bounds\\ \eqref{eq.uricte}, \eqref{eq.lricte}\end{tabular}} & \multicolumn{1}{c}{\cellcolor[HTML]{FFFEFC}\begin{tabular}[c]{@{}c@{}}\ac{TW}\vspace{-2mm}\\\vspace{-1.5mm}  approximation\\ \eqref{eq.approxuricte}, \eqref{lricte2}\end{tabular}} & \cellcolor[HTML]{FFFEFC}\begin{tabular}[c]{@{}c@{}}Empirical \vspace{-2mm}\\\vspace{-1.5mm}lower bounds\\\cite{DosPeyFad:10}\end{tabular} 
		&\cellcolor[HTML]{FCFEFC}  \begin{tabular}[c]{@{}c@{}} BCT \\\vspace{2mm}  \cite{BlaCarTan:11}\end{tabular} 
		& \cellcolor[HTML]{FCFEFC} \begin{tabular}[c]{@{}c@{}}BT\\\vspace{2mm} \cite{BahTan:10}\end{tabular} \\ \hline
		\cellcolor[HTML]{FAF8EA}                        & \multicolumn{1}{|c}{\cellcolor[HTML]{FCF5E4}$0.3071$}                                                  & \multicolumn{1}{c}{\cellcolor[HTML]{FCF5E4}$0.3395$}                                                         & \cellcolor[HTML]{FCF5E4}$0.2703$                                  &  \cellcolor[HTML]{E9F6E3}$0.3408$                                        & \cellcolor[HTML]{E9F6E3}$0.3402$                        \\ \cline{2-6} 
		\multirow{-2}{*}{\cellcolor[HTML]{FAF8EA}$0.4$} & \multicolumn{1}{|c}{\cellcolor[HTML]{FFFEFC}$0.2561$}                                                  & \multicolumn{1}{c}{\cellcolor[HTML]{FFFEFC}$0.2846$}                                                         & \cellcolor[HTML]{FFFEFC}$0.2322$                                  &  \cellcolor[HTML]{FCFEFC}$0.2777$                                        & \cellcolor[HTML]{FCFEFC}$0.2772$                        \\ \hline
		\cellcolor[HTML]{FAF8EA}                        & \multicolumn{1}{|c}{\cellcolor[HTML]{FCF5E4}$0.3000$}                                                  & \multicolumn{1}{c}{\cellcolor[HTML]{FCF5E4}$0.3304$}                                                         & \cellcolor[HTML]{FCF5E4}$0.2626$                                  &  \cellcolor[HTML]{E9F6E3}$0.3344$                                        & \cellcolor[HTML]{E9F6E3}$0.3337$                        \\ \cline{2-6} 
		\multirow{-2}{*}{\cellcolor[HTML]{FAF8EA}$0.6$} & \multicolumn{1}{|c}{\cellcolor[HTML]{FFFEFC}$0.2512$}                                                  & \multicolumn{1}{c}{\cellcolor[HTML]{FFFEFC}$0.2778$}                                                         & \cellcolor[HTML]{FFFEFC}$0.2268$                                  &  \cellcolor[HTML]{FCFEFC}$0.2734$                                        & \cellcolor[HTML]{FCFEFC}$0.2729$                        \\ \hline
		\cellcolor[HTML]{FAF8EA}                        & \multicolumn{1}{|c}{\cellcolor[HTML]{FCF5E4}$0.2949$}                                                  & \multicolumn{1}{c}{\cellcolor[HTML]{FCF5E4}$0.3239$}                                                         & \cellcolor[HTML]{FCF5E4}$0.2580$                                  &  \cellcolor[HTML]{E9F6E3}$0.3297$                                        & \cellcolor[HTML]{E9F6E3}$0.3291$                        \\ \cline{2-6} 
		\multirow{-2}{*}{\cellcolor[HTML]{FAF8EA}$0.8$} & \multicolumn{1}{|c}{\cellcolor[HTML]{FFFEFC}$0.2477$}                                                  & \multicolumn{1}{c}{\cellcolor[HTML]{FFFEFC}$0.2729$}                                                         & \cellcolor[HTML]{FFFEFC}$0.2214$                                  &  \multicolumn{1}{c}{\cellcolor[HTML]{FCFEFC}$0.2703$}                   & \multicolumn{1}{c|}{\cellcolor[HTML]{FCFEFC}$0.2698$}   \\ \hline
	\end{tabular}
\end{table*}
%


With the aim of comparing different sufficient recovery conditions  via $\ell_1$-minimization,  \ac{iht}, and \ac{cosamp} algorithms, in Fig. \ref{fig.Mor_deltas_third} we report  the maximum oversampling ratio, $\maxs/m$, such that $\ppr \geq 0.999$. All curves have been obtained by using the \ac{EED} based approach. Specifically, for $\ell_1$-minimization we consider the symmetric \ac{ric} condition $\ric\leq 1/3$~\cite{CaiTonZha:13}, its relaxed asymmetric extension $\mu_\text{\tiny{ECG}}(s,{\bf A}) < 1$ proposed in Section~\ref{sec.apr}, $\delta_{2s}({\bf A})<0.624$~\cite{FouRau:13}, $\mu_\text{\tiny{FL}}(s,{\bf A}) <1$~\cite{FouLai:08}, and $\mu_\text{\tiny{BT}}(s,{\bf A})<1$~\cite{BlaTho:10}. {For \ac{iht} we used the conditions $\delta_{3s}({\bf A})<0.5773$~\cite{Fou:11} and $\mu_\text{\tiny{BCTT}}(s,{\bf A})<1$~\cite{BlaCarTanThom:11}}, while for the  \ac{cosamp} we considered $\delta_{4s}({\bf A})<0.4782$~\cite{FouRau:13}. We can see that the asymmetric conditions provide higher estimates of the sparsity which can be handled by compressed sensing, compared to the symmetric conditions (more than $40\%$ increase in $s$). 
As known, the $\ell_l$-minimization  and \ac{iht} algorithms allow higher oversampling ratios than the \ac{cosamp} algorithm.  
\begin{figure}[t]
	\centering
	\psfrag{m/n}{\scriptsize $m/n$}
	\psfrag{s/m}{\scriptsize $\maxs/m$}
	\psfrag{L2s+(L6s+U6s)/4}{\scriptsize $\ell_1$, $\mu_\text{\tiny{BT}}(s,{\bf A}) <1$}
	\psfrag{delta2s<.624}{\scriptsize $\ell_1$, $\delta_{2s}({\bf A})<0.624$}
	\psfrag{11L(12s)+U(11s)<10}{\scriptsize $\ell_1$,  $11\,\underline{\delta}_{12\,s}({\bf A})+\overline{\delta}_{11\,s}({\bf A}) <10$}
	\psfrag{mufl(s,A)<1}{\scriptsize  $\ell_1$, $\mu_\text{\tiny{FL}}(s,{\bf A}) <1$}
	\psfrag{2L(s)+U(s)<1}{\scriptsize $\ell_1$, $\mu_\text{\tiny{ECG}}(s,{\bf A})<1$}
	\psfrag{deltas<1/3}{\scriptsize $\ell_1$, $\ric <1/3$}
	\psfrag{IHT}{\scriptsize \ac{iht}, $\delta_{3s}({\bf A})<0.5773$}
	\psfrag{CoSaMP}{\scriptsize \acs{cosamp}, $\delta_{4s}({\bf A})\!<\!0.4782$}
	\psfrag{IHTmuBCTT(s,A)<1}{\scriptsize \acs{iht}, $\mu_\text{\tiny{BCTT}}(s,{\bf A})<1$}
	\psfrag{e-3}{\scriptsize$ \times 10^{-3}$}
	\includegraphics[clip,width=0.99\linewidth,height=0.77\linewidth]{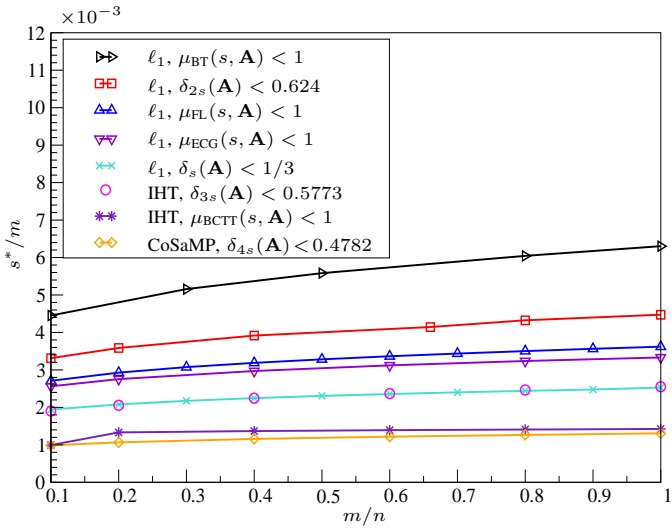}
	\caption{The maximum oversampling ratio, $\maxs/m$, for various recovery algorithms and their associated sufficient conditions using the proposed \ac{EED} based approach, for $m=4000$ and $\ppr \geq 0.999$ ($\eta=10^{-3}$).}
	\label{fig.Mor_deltas_third}
\end{figure}
\begin{figure}[t]
	\centering
	\psfrag{m/n}{\scriptsize $m/n$}
	\psfrag{s/m}{\scriptsize $\maxs/m$}
	\psfrag{L2s+(L6s+U6s)/4}{\scriptsize  RIP analysis (\ac{EED}), $\mu_\text{\tiny{BT}}(s,{\bf A}) <1$}
	\psfrag{Donohofinite}{\scriptsize   Polytope  analysis \cite{DonTan:10c}}
	\psfrag{Geometric func analysis  [6]}{\scriptsize  Geometric functional  analysis \cite{RudRom:08}}
	\psfrag{Null space}{\scriptsize Null space property \cite{FouRau:13} }
	\psfrag{RIP Analysis using CS Book analysis delta2s<0.6147}{\scriptsize RIP analysis \cite{FouRau:13}, $\delta_{2s}({\bf A})<0.624$}
	\psfrag{e-3}{\scriptsize$ \times 10^{-3}$}
	\includegraphics[clip,width=0.99\linewidth,height=0.77\linewidth]{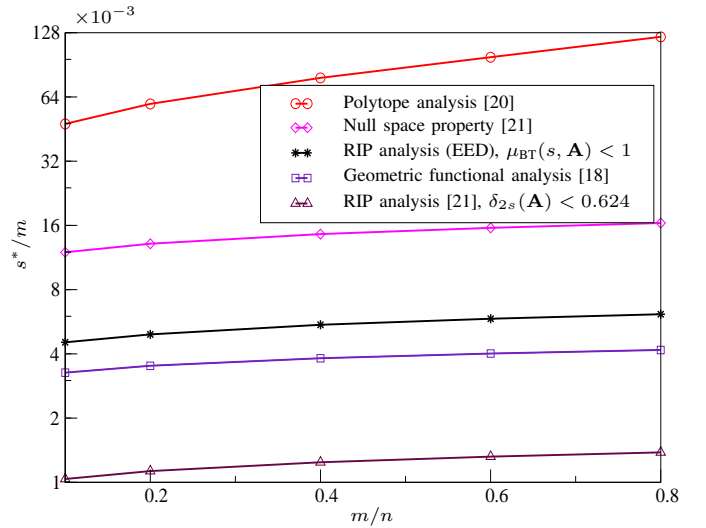}
	\caption{The maximum oversampling ratio, $\maxs/m$, for perfect recovery via  $\ell_1$-minimization, estimated by the proposed \acs{rip} based approach (\ac{EED}) along with the \ac{rip} \cite{FouRau:13},  polytope \cite{DonTan:10c}, Null space \cite{FouRau:13}, and geometric functional \cite{RudRom:08} analyses, for $m=4000$ and $\ppr \geq 0.5$.}
	\label{fig.Mor_deltas_Don}
\end{figure}
\begin{figure}[t]
	\psfrag{s/m}{\scriptsize $s/m$}
	\psfrag{m/n}{\scriptsize $m/n$}
	\psfrag{e-4}{\scriptsize $\times 10^{-4}$}
	\vspace{-0.1cm}
	\includegraphics[clip,width=0.99\linewidth,height=0.77\linewidth]{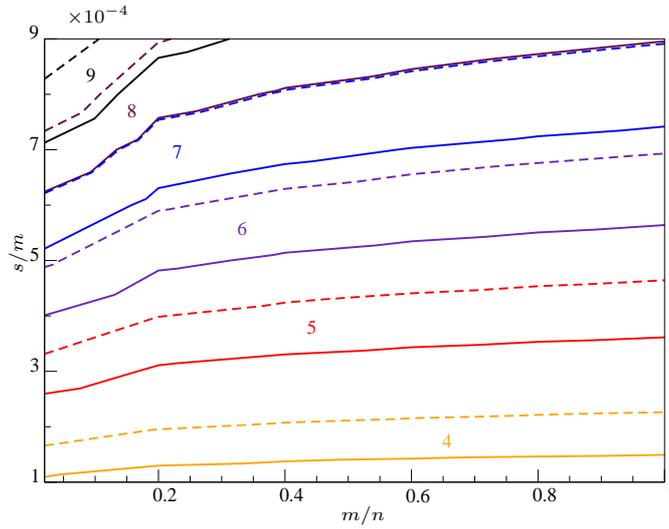}\centering
	\caption{Level sets of robustness and stability thresholds in Section \ref{sec.robustness}, $C_{1}^{*}$ (solid) and $C_{2}^{*}$ (dashed),  with $C_{1}^{*}, \,C_{2}^{*}$ $\in \{4,5,6,7,8,9\}$, for $m=2\cdot10^{4}$ and $\epsilon_{1}=\epsilon_{2}=10^{-3}$, using the \ac{EED} based approach.}
	\label{fig.c1c2}
\end{figure}
%
 
Moreover, we provide in  Fig. \ref{fig.Mor_deltas_Don}  the maximum oversampling ratio, for uniform recovery, indicated by our proposed approach along with those obtained from the  polytope \cite{DonTan:10c}, Null space \cite[Theorem $9.29$]{FouRau:13}, geometric functional \cite[Theorem $4.1$]{RudRom:08}, and \ac{rip} \cite[Theorem $9.27$]{FouRau:13} analyses for finite matrices with $m=4000$ and $\ppr\geq 0.5$. However, we would like to note that the polytope based approach suggests tighter bounds on the maximum sparsity order, as it fully exploits the geometry of the  $\ell_1$-minimization for signal recovery from Gaussian measurements. On the other hand, the RIP is suitable for analyzing the robust and stable reconstruction with several sparse recovery algorithms, such as optimization, greedy, and thresholding. 

Finally, regarding the analysis for compressible signals in noise, the contours for robustness and stability  thresholds $C_{1}^{*}$ and $C_{2}^{*}$ are shown in Fig. \ref{fig.c1c2}. As can be seen for small $s/m$ the thresholds are small, indicating that the more sparse is the signal, the more robust and stable is the reconstruction process. Therefore, a compromise between sparsity and robustness/stability should be considered when designing the acquisition system. This figure also gives the maximum oversampling ratio for a given $m$ and $n$, such that the minimization program \eqref{eq.bpdn} can approximately recover the measured signal with a predefined discrepancy. 
 \section{Conclusion}
 \label{sec.conclusion}
For sparse data acquisition we have found that the concentration of measure inequality provides a loose upper bound on the probability that a measurement submatrix is ill conditioned.  For example, in some cases it overestimate the maximum sparsity ratio by over $220\%$ with respect to the proposed exact eigenvalues based approach.  
  For finite matrices, by tightly bounding the symmetric and asymmetric \acp{ric}, the best current lower bound on the maximum sparsity order guaranteeing successful recovery has been provided, for various sparse reconstruction algorithms. 
  For stable and robust recovery of compressible data, we have noticed that when the sparsity order decreases the discrepancy between the recovered and original signals reduces. Finally, we have shown that simple approximations for the \acp{ric} can be obtained based on \ac{TW} distributions.
  

\bibliographystyle{IEEEtran}
\bibliography{IEEEabrv,Nov_2018_ArXiv.bib}
\pagebreak
\begin{IEEEbiography}{Ahmed~Elzanaty}
	(S'13) received the B.Sc. (\textit{with honors}) and  M.Sc. degrees in Electronics and Communications Engineering from Port Said University, Egypt, in 2008 and 2013, respectively, and the Ph.D. degree (\textit{excellent cum laude}) in Electronics,  Telecommunications, and Information technology from the University of Bologna, Italy, in 2018. He was a recipient of a doctoral scholarship from the EU-METALIC II project, within the framework of Erasmus Mundus Action 2. Currently, he is a research fellow at the University of Bologna. He has participated in several national and European projects, such as GRETA and EuroCPS. His research interests include statistical signal processing and digital communications, with particular emphasis on compressed sensing and sparse source coding. He was the recipient of the best paper award at the IEEE Int. Conf. on Ubiquitous Wireless Broadband (ICUWB 2017).  Dr. Elzanaty was a member of the Technical Program Committee of the European Signal Processing Conf. (EUSIPCO 2017 and 2018). He is also a representative of the IEEE Communications Society's Radio
	Communications Technical Committee for several international conferences. 
\end{IEEEbiography}	
\begin{IEEEbiography}{Andrea~Giorgetti}
	(S'98--M'04--SM'13) received the Dr. Ing. degree (\textit{summa cum
		laude}) in electronic
	engineering and the Ph.D. degree in electronic engineering and
	computer science from the University of Bologna, Italy, in 1999 and
	2003, respectively.
	From 2003 to 2005, he was a Researcher with the National Research
	Council, Italy. He joined the Department of Electrical, Electronic,
	and Information Engineering ``Guglielmo Marconi,'' University of
	Bologna, as an Assistant Professor in 2006 and was promoted to
	Associate Professor in 2014.
	In spring 2006, he was with the Laboratory for Information and
	Decision Systems (LIDS), Massachusetts Institute of Technology (MIT),
	Cambridge, MA, USA. Since then, he has
	been a frequent visitor to the Wireless Information and Network
	Sciences Laboratory at the MIT, where he presently holds the Research
	Affiliate appointment. His research interests include ultrawide bandwidth communication systems, active and passive localization, wireless sensor networks,
	and cognitive radio. He has co-authored the book \textit{Cognitive Radio Techniques: Spectrum Sensing, Interference Mitigation, and Localization} (Artech House, 2012). He was the Technical Program Co-Chair of several symposia at the IEEE Int. Conf. on Commun. (ICC), and IEEE Global Commun. Conf. (Globecom). He has been an Editor for the \textsc{IEEE Communications Letters} and for the \textsc{IEEE Transactions on Wireless Communications}. He has been elected Chair of the IEEE Communications Society's Radio Communications Technical Committee.
\end{IEEEbiography}
\begin{IEEEbiography}{Marco~Chiani}
	(M'94--SM'02--F'11) received the Dr. Ing. degree (\textit{summa cum laude}) in electronic engineering and the Ph.D. degree in electronic and computer engineering from the University of Bologna, Italy, in 1989 and 1993, respectively. He is a Full Professor in Telecommunications at the University of Bologna. During summer 2001, he was a Visiting Scientist at AT\&T Research Laboratories, Middletown, NJ. Since 2003 he has been a frequent visitor at the Massachusetts Institute of Technology (MIT), Cambridge, where he presently holds a Research Affiliate appointment. His research interests are in the areas of communications theory, wireless systems, and statistical signal processing, including MIMO statistical analysis, codes on graphs, wireless multimedia, cognitive radio techniques, and ultra-wideband radios. In 2012 he has been appointed Distinguished Visiting Fellow of the Royal Academy of Engineering, UK. He is the past chair (2002--2004) of the Radio Communications Committee of the IEEE Communication Society and past Editor of Wireless Communication (2000--2007) for the journal \textsc{IEEE Transactions on Communications}. He received the 2011 IEEE Communications Society Leonard G. Abraham Prize in the Field of Communications Systems, the 2012 IEEE Communications Society Fred W. Ellersick Prize, and the 2012 IEEE Communications Society Stephen O. Rice Prize in the Field of Communications Theory.
\end{IEEEbiography}

\end{document}